\input{aipcheck}

\documentclass[
    ,final           
  ]
  {aipproc}

\layoutstyle{6x9}

\begin{document}

\title{Dynamos and Chemical Mixing in Evolved Stars }


\author{J. Nordhaus\footnote{nordhaus@pas.rochester.edu} \space and E. G. Blackman}{
  address={Dept. of Physics and Astronomy, University of Rochester, Rochester, NY 14627}
,altaddress={Laboratory for Laser Energetics, University of Rochester, Rochester, NY 14623}}

\begin{abstract}
In low-mass Red Giant Branch (RGB) and Asymptotic Giant Branch (AGB) stars, anomalous mixing must transport material near the hydrogen-burning shell to the convective envelope.  Recently, it was suggested that buoyant magnetic flux tubes could supply the necessary transport rate \cite{2007arXiv0708.2949B}.  The fields are assumed to originate from a dynamo operating in the stellar interior.  Here, we show what is required of an $\alpha-\Omega$ dynamo in the envelope of an AGB star to maintain these fields.  Differential rotation and rotation drain via turbulent dissipation and Poynting flux, so if shear can be resupplied by convection, then large-scale toroidal field strengths of $\left<B_\phi\right>\simeq3\times10^4$ G can be sustained at the base of the convection zone.  

\end{abstract}


\maketitle


\section{Introduction}

In low-mass Red Giant Branch (RGB) and Asymptotic Giant Branch (AGB) stars, an anomalous transport mechanism must cycle material from the convective envelope to the stable radiative region where nuclear processing can occur \cite{1991ApJ...371..578G,1995ApJ...447L..37W}.   In particular, material must be pumped downward from the convection zone, exposed to partial H burning (so-called Cool Bottom Processing; CBP) and returned to the convective envelope.  Isotopic abundances provide constraints on the required processing temperatures and mass transfer rates.  Recently, it was proposed that buoyant magnetic flux tubes can provide the necessary transport \cite{2007arXiv0708.2949B}.

In evolved stars, magnetically mediated outflows have been proposed as the origin of bipolarity in post-Asympotitc Giant Branch stars (post-AGB) and planetary nebula (PN) \cite{1997ApJ...489..946P,2001Natur.409..485B,2006MNRAS.370.2004N,2007MNRAS.376..599N}.  The needed large-scale field likely results from a dynamo operating during the RGB or AGB.  In both phases, a convective envelope coupled with differential rotation in the stellar interior can amplify the field via an $\alpha-\Omega$ dynamo \cite{2007MNRAS.376..599N}.  However, as the field amplifies, differential rotation energy is drained and rapidly terminates the dynamo ($\leq 100$ yrs) \cite{1997ApJ...489..946P,2001Natur.409..485B}.  This would favor binary shaping mechanisms for post-AGB/PNe and a binary induced dynamo (see \cite{2007MNRAS.376..599N,2007arXiv0707.3792N} for detail).  Only if convection resupplies differential rotation (analogous to the $\lambda$-effect in the Sun \cite{2004muga.book.....R}) could a sustained dynamo operate throughout the RGB or AGB phases without a binary \cite{2007MNRAS.376..599N}.  While sustained fields are required to shape post-AGB outflows, the dynamo required for mixing may be weaker and need not influence the outflow.  It may be that a weaker dynamo (supplying sufficient mixing) is sustained throughout the AGB phase, but that a binary companion is required to power post-AGB bipolarity.

Here, we present a dynamical $\alpha-\Omega$ dynamo operating at the base of the convection zone in a 3 $M_\odot$ AGB star.  We consider two models: (i.) shear is not resupplied, and a transient dynamo results (ii.) convection resupplies shear, sustaining the dynamo.  For both models, the field penetrates the shear zone and may transport material to the H burning shell.  We compare our results to the field strengths and penetration depths needed for viable magnetic mixing in low-mass RGB and AGB stars \cite{2007arXiv0708.2949B}.  We comment on future directions.

\begin{figure}
  \includegraphics[height=.28\textheight]{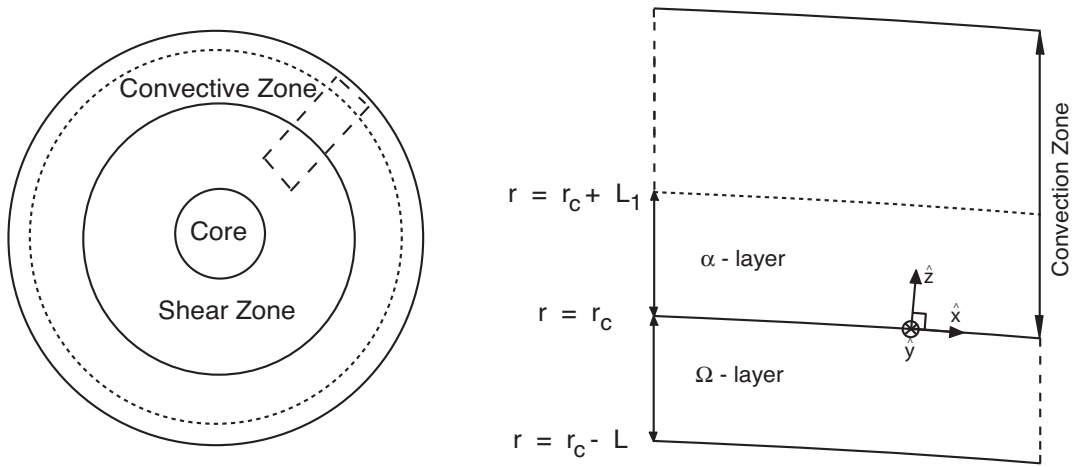}
  \caption{Figure from \cite{2007MNRAS.376..599N}.  A meridional slice of dynamo geometry.  The left figure shows the global geometry of the AGB star.  The right figure is a close-up view of the dashed region on the left.  The $\alpha$-effect is driven by convection and occurs in layer of thickness $L_1$ above the differential rotation zone.  The poloidal component of the  field is pumped downwards into the differential zone, where it is wrapped torodially due to the $\Omega$-effect.}
  \label{geometry}
\end{figure}

\section{An Evolved Star Dynamo}

A 2-D schematic of our 1-D, 3 $M_\odot$ AGB star\footnote{Stellar model courtesy of S. Kawaler -- personal communcation.} is presented in Fig. \ref{geometry}.  In such a star, the convection zone extends from the stellar surface to the interface between the convective and radiative zones.  In this layer, convective twisting motions convert buoyant toroidal field into poloidal field through the $\alpha$-effect.  Below the convection layer, the differential rotation zone extends to the core and shears poloidal field back into toroidal field through the $\Omega$-effect.

The differential rotation profile in the AGB interior is unknown.  In order to employ a reasonable initial profile, we assume that the main sequence surface rotation velocity is 200 km s$^{-1}$.  A radial shear profile is then obtained by conserving angular momentum on spherical shells during post-MS evolution \cite{2006MNRAS.370.2004N}.  

We solve for the time evolution of dynamic quantities at the interface between the convective region and shear zone ($r=r_c$).  The rotation profile across the shear layer varies from $\Omega$ at the interface to $\Omega+\Delta\Omega$ at $r=r_c-L$.  Thus, $\Delta\Omega$ is a measure of shear in the differential rotation zone.  If $\Delta\Omega=0$, the system exhibits solid body rotation.  The average poloidal field, $\left<B_p\right>$, and average toroidal field, $\left<B_\phi\right>$ amplify from 1 G seed values until they are quenched through a drain of the available differential rotation energy.  We refer to reader to \cite{2007MNRAS.376..599N} for a derivation of the precise 1-D, mean-field equations solved.

\subsection{Shear Zone Penetration Depth}
To capture aspects of 2-D, mean-field geometry within the framework of our simple, 1-D time-dependent model, we employ two turbulent diffusion coefficients: $\beta_p$, corresponding to diffusion of the poloidal field (which grows primarily in the convective region) and $\beta_\phi$, corresponding to diffusion of the toroidal field (which is amplified in the differential rotation zone; see Fig. 1).  We also use $\beta_\phi$ as the turbulent diffusion coefficient for the toroidal velocity.  Since the convective region is highly turbulent and the differential rotation zone is more weakly turbulent, we take $\beta_\phi\ll\beta_p$.  The value of $\beta_\phi$ governs how far the poloidal component of the field can diffuse into the shear zone in a cycle period.  The further into the shear zone the poloidal field can penetrate, the greater the shear energy that can be extracted and utilized by the dynamo.  The coefficient, $\beta_\phi$ also governs how much heat is generated by turbulent dissipation in the shear layer.  

To quantitfy this, we define $\delta$ as the depth at which the poloidal field can diffuse into the shear layer in a cycle period, $\tau$.  We have

\begin{equation}
\delta\simeq(\beta_\phi\tau)^{\frac{1}{2}}.
\end{equation}

\noindent If $\delta/L=1$, the poloidal component of the field reaches the core boundary and extracts all available shear energy.  If $\delta/L<1$, then the poloidal field extracts a fraction of the total shear energy and can not penetrate as deeply.  Thus for a given $\beta_\phi$, we can determine the depth to which the field diffuses.  As an aside, we note that for the maximally growing dynamo mode, $\tau$ is weakly dependent on $\beta_\phi$ and hence, would change the power index slightly.  Although we do not expect this effect to be significant, future work should use the maximally growing mode.


\begin{figure}[h!]
  \includegraphics[height=.43\textheight]{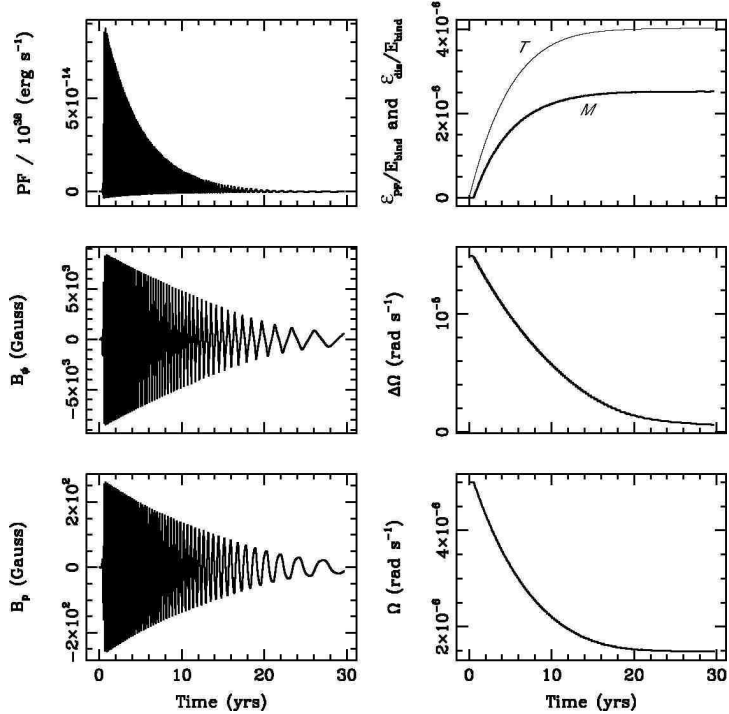}
  \caption{Figure from \cite{2007MNRAS.376..599N}.  The differential rotation energy is allowed to drain through field amplification and turbulent dissipation.  We define, $\epsilon_{\left[PF,dis\right]}\equiv\int^t_0E_{\left[PF,dis\right]}\left(t'\right)dt'$ and label $M$ ($\epsilon_{PF}$) and $T$ ($\epsilon_{dis}$) on the top right plot to distinguish between the thermal and magnetic contributions.  If $\epsilon_{[PF,dis]}/E_{bind}\geq1$, then the AGB envelope was unbound via heat from turbulent dissipation or time-integrated Poynting flux.  Peak field strengths are a factor of $\sim5-10$ less then those obtained in (\cite{2001Natur.409..485B}).  Differential rotation energy is drained in $<20$ yrs.  Lowering the turbulent dissipation coefficient in the shear region results in the differential rotation energy draining at a slower rate, allowing the field to sustain for longer periods of time ($\sim40-50$ yrs).  However, peak field strengths remain the same and are a function of $\Delta\Omega$ and $\Omega$.}
  \label{2}
\end{figure}

\subsection{The Need For Convection Resupplying Shear}
To investigate the back-reaction of field growth on the shear, we apply our dynamical model using the initial rotation profile previously described.  Our results are shown in Fig. \ref{2}.  The dynamo terminates rapidly as the shear energy is drained after $\sim20$ yrs.  In addition, the loss of Poynting flux spins down the interface by a factor of $\sim3$.  In order to sustain the dynamo through an AGB lifetime, a constant differential rotation profile must be established.  This occurs in the sun as convection re-seeds shear through the $\lambda$-effect \cite{2004muga.book.....R}.  Although it remains to be established if a similar effect occurs in evolved stars, by analogy to the solar case, we allow a fraction of the turbulent energy cascade to resupply shear.  Additionally, we keep the rotation at the interface fixed.  This is physically equivalent to storing Poynting flux in the interface region.  If the field is trapped, Poynting flux does not emerge from the layer and thus, does not spin down the envelope.  An AGB dynamo will thus sustain when the following two conditions are met: (i.) convection resupplies shear (ii.) Poynting flux is stored inside the envelope \cite{2007MNRAS.376..599N}.

For our $3$ $M_\odot$ AGB star, under those conditions, a steady-state dynamo is established when $\sim1\%$ of the turbulent cascade energy per unit time reinforces the shear (see Fig. \ref{3}).

\section{Requirements for Magnetic Mixing}

\begin{figure}[h!]
  \includegraphics[height=.43\textheight]{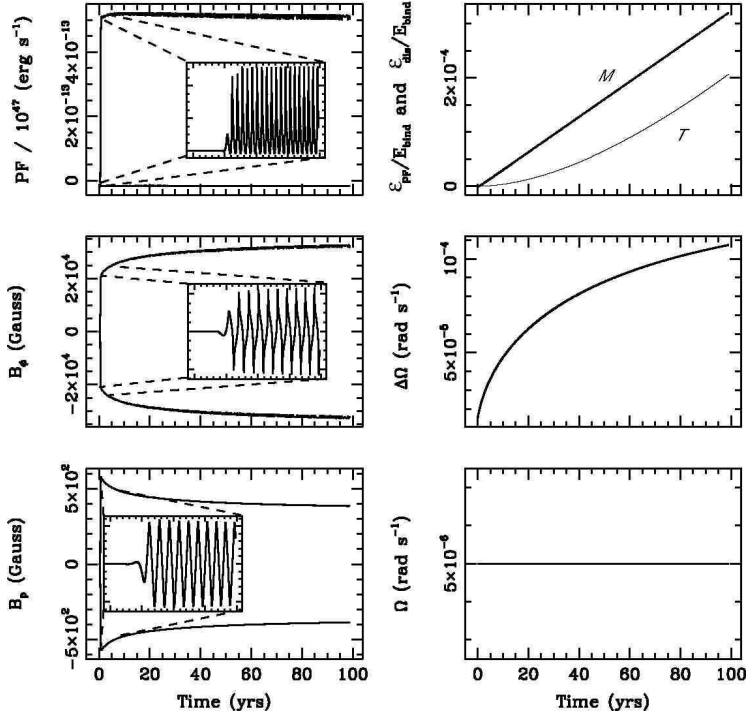}
  \caption{Figure from \cite{2007MNRAS.376..599N}.  Convective resupply results in a steady-state differential rotation profile.  For the left column, the envelope of the poloidal, toroidal and Poynting flux is plotted.  The Poynting flux is sustained at $\sim5\times10^{34}$ erg/s.  The sustained Poynting flux supplies enough energy to unbind the envelope of our $3$ $M_\odot$ model at the end of the AGB phase ($\sim10^5$ yrs).  In this figure, $\sim0.1\%$ of the cascade power must be used to sustain the  differential rotation to supply the requisite Poynting flux.  This choice of parameters corresponds to a magnetically dominated explosion at the end of the AGB.}
  \label{3}
\end{figure}

For a given bulk transport rate, Cool Bottom Processing predicts the chemical and isotopic evolution of the convective envelope.  Without appealing to a specific driving mechanism, CBP yields explicit temperature requirements and nuclear processing times needed to match observations for a specific stellar model.  To investigate the physical origin of CBP, hydrodynamical models, such as rotationally induced meridional mixing were suggested.   However, these models do not supply sufficient mixing to explain observations of low-mass RGB field and globular cluster stars \cite{2006A&A...453..261P}.

Recently, as an alternative to hydrodynamic models, magnetically induced mixing was proposed \cite{2007arXiv0708.2949B}.  Buoyant toroidal fields in the radiative zone carry material to the convective envelope and drive mass circulation.   By requiring CBP transport rates, it is possible to constrain magnetic field strengths and buoyant velocities necessary for a given stellar model.  

For an initial 1.5 $M_\odot$ star with metallicity one-half solar \cite{1997ApJ...478..332S}, constraints are calculated for the subsequent RGB and AGB phases.  In both phases, subadiabatic zones similar to the solar tachocline exist \cite{2003ApJ...582.1036N}.  Since our previous dynamo calculations apply to an AGB star, we focus on that phase.

For the AGB phase, the total mass has decreased to $M=1.2$ $M_\odot$.  The base of the convection zone is located at $r_c=5.7\times10^{10}$ cm.  For a mass flux of $\dot{M}=10^{-6}$ $M_\odot/yr$, material must mix to a radius of $r_{CBP}=2\times10^9$ cm which is slightly above the location of the H burning shell ($r_H=1.5\times10^9$ cm) \cite{2007arXiv0708.2949B}.  

Given a mass transfer rate, mixing depth and buoyant rise velocity, the magnetic field strength at $r_H$ and $r_c$ can be constrained.  If the field is purely toroidal and in flux tubes that rise individually, then $B_t=B\left(r_{CBP}\right)=5\times10^6$ G \cite{2007arXiv0708.2949B}.  At the base of the convection zone, this corresponds to a field of $B_c= B\left(r_{c}\right)\simeq9\times10^3$ G for a dipole geometry.

Direct comparison to the dynamo calculation is difficult as our AGB star is 2.5 times more massive.  In addition, our $3$ $M_\odot$ structure corresponds to a star that has just reached the AGB.  Future work should investigate a dynamo operating in a lower mass star, during the interpulse periods.  Nevertheless, for plausibility arguments, a general comparison with our sustained dynamo is useful.  

From our 3 $M_\odot$ AGB star, the toroidal field strength at the base of the convection zone is $\left<B_{\phi,c}\right>\sim3\times10^4$ G, suggesting that it may be possible to produce sufficiently high average fields.  However, since we do not solve for the radial structure, we cannot determine the field strength where CBP nucleosynthesis occurs.  

We can, however, estimate the depth to which the magnetic field penetrates the shear layer.  Based on the initial conditions used in Fig. \ref{3}, Eqn. 1 gives that the field reaches a radius of $\sim3\times10^{10}$ cm with approximately 15 cycles per year.  This is encouraging and suggests that significant penetration into the shear zone may occur.  

\subsection{Flux Tubes versus Average Fields}
Our dynamo calculations are mean-field and correspond to 
volume averages. When the ratio of average thermal to magnetic pressure
 $\beta={\frac{\left< P\right>}{ \left< B^2\right>/8\pi}}>>1$ it is possible for the magnetic field to be concentrated in flux tubes with an overall
 volume filling fraction of the magnetic field  $f<<1$, to be small, such that
$f\sim \frac{1}{ \beta}$ \cite{1996PhRvL..77.2694B}.  
When flux tubes are in presure balance with their exterior, 
 the mean-field strength is then related to a flux tube field, $B_t$ 
 by $\left<B^2\right> \sim f\left< B_t^2\right>$ \cite{1996PhRvL..77.2694B}.  
Since the field constraints on magnetic mixing from Ref. \cite{2007arXiv0708.2949B} constrain the magnitude of flux tube fields with total $f<<1$, 
the mean field can be much lower than the flux tube field, relaxing the demands
on the strength of the dynamo.

Further investigation  of all of the above is warranted for relevant stellar models.  Determining the radial diffusion coefficients will constrain the amount of shear energy available to the dynamo and thus, determine the penetration depth and field strengths.  For a given stellar model, constraints on the fraction of convective energy needed to sustain a steady shear profile should be calculated in both the RGB and AGB phases.  In the RGB phase, weaker fields are required.  Realistic rotation profiles should be utilized when known.  Finally, $\geq$ 2-D model calculations incorporating $\Omega$-quenching will provide radial information on magnetic field strengths.

\section{Conclusions}


It was recently suggested that buoyant toroidal fields, amplified in the stellar interior, may supply the necessary chemical transport rates from the hydrogen burning shell to the convective envelope required from Cool Bottom Processing in low-mass RGB and AGB stars \cite{2007arXiv0708.2949B}. 

In this paper, we have presented results of a dynamic $\alpha-\Omega$ dynamo operating in an AGB star.  Amplification of the large-scale field drains shear energy rapidly ($\leq100$ yrs.).  For an isolated star, stringent conditions (re-supply of shear + storage of Poynting flux) must be met if a dynamo is to sustain through the RGB or AGB phases.  This can be accomplished if $\sim1\%$ of the turbulent energy cascade reinforces shear.  In this case, our AGB model produces strong toroidal fields ($B_\phi\simeq3\times10^4$ G) at the base of the convection zone that can diffuse into the shear layer.  This suggests that magnetic fields may be a plausible mechanism for mixing in low-mass RGB and AGB stars, however  future work is needed before this can be firmly established.



\begin{theacknowledgments}
  JN thanks M. Busso and G. J. Wasserburg for their gracious invitation and ensuing discussions.  This research was carried out under the financial support of a Horton Fellowship from the Laboratory for Laser Energetics through the U. S. Department of Energy and HST grant AR-10972.
\end{theacknowledgments}



\bibliographystyle{aipproc}   

\bibliography{sample}

\IfFileExists{\jobname.bbl}{}
 {\typeout{}
  \typeout{******************************************}
  \typeout{** Please run "bibtex \jobname" to optain}
  \typeout{** the bibliography and then re-run LaTeX}
  \typeout{** twice to fix the references!}
  \typeout{******************************************}
  \typeout{}
 }

\end{document}